\documentclass[aps,twocolumn,showpacs]{revtex4}
\usepackage{graphicx}
\usepackage{dcolumn}
\usepackage{bm}

\begin{document}

\title{Crises in a dissipative Bouncing ball model}

\author{Andr\'e L.\ P.\ Livorati$^{1,2}$, Iber\^e L. Caldas$^1$, Carl P. Dettmann$^{2}$ and Edson D.\ Leonel$^{3}$}

\affiliation{$^1$Instituto de F\'isica - IFUSP - Universidade de S\~ao
Paulo - USP  Rua do Mat\~ao, Tr.R 187 - Cidade Universit\'aria --
05314-970 -- S\~ao Paulo -- SP -- Brazil\\
$^2$ School of Mathematics, University of Bristol, Bristol, BS8 1TW, United Kingdom\\
$^3$ Departamento de F\'isica -- UNESP -- Univ Estadual Paulista -- Av. 24A, 1515 - Bela Vista -
13506-900 - Rio Claro - SP - Brazil
\\ 
}

\begin{abstract}
The dynamics of a bouncing ball model under the influence of dissipation is investigated by using a two dimensional nonlinear mapping.
When high dissipation is considered, the dynamics evolves to different attractors.
The evolution of the basins of the attracting fixed points is characterized, as we vary the control
parameters. Crises between the attractors and their boundaries are observed.
We found that the multiple attractors are intertwined, and when the boundary crisis between their stable and unstable manifolds occur,
it creates a successive mechanism of destruction for all attractors originated by the sinks. Also, an impact physical crises is setup,
and it may be useful as a mechanism to reduce the number of attractors in the system.
\end{abstract}

\maketitle

\section{Introduction}
\label{sec1}
~~Modelling of dynamical systems is one of the most embracing area of interest among physicists and 
mathematicians in general \cite{ref1}. Very popular among these models are systems that are low-dimensional dynamics \cite{ref2,ref3}, which besides
the simple modelling, such systems can present a very complex dynamics leading to a rich variety of nonlinear phenomena 
\cite{ref3,ref4,ref5,ref6}, including bifurcations in nonsmooth dynamical systems \cite{add1}.  

Here we study the problem of a bouncing ball model, where a free particle is suffering collisions with a vibrating 
wall under the presence of a constant gravitational field. Holmes \cite{ref7,ref8} and Pustylnikov \cite{ref9,ref10} were among the first to
study the bouncing ball dynamics. This model has been used in many physical and engineering applications. For instance, it describes
a similar acceleration phenomenon that cosmic rays experiences to acquire high energies \cite{ref11}; the dynamic stability in human
performance, where a human tries to stabilise a ball on a vibrating tennis racquet \cite{ref12}; and the subharmonic vibrations waves
in a nanometre-sized mechanical contact system \cite{ref13}. One can also find studies in granular materials \cite{ref14,ref15,ref16,ref17}, experimental
devices concerning normal coefficient of restitution \cite{ref18,ref19}, mechanical vibrations \cite{ref20,ref21,ref22}, anomalous transport
and diffusion \cite{ref23,ref24}, thermodynamics \cite{ref25}, chaos control \cite{ref26,ref27,ref28}, besides the well known connection
with the standard mapping \cite{ref2}, which leads to other several applications. 

Although the bouncing ball 
problem has been studied for many years \cite{ref7,ref8,ref9,ref10,ref29,ref30}, concerning different aspects and applications, 
the implications of the nonlinear perturbation requires an extensive and complex analysis where some chaotic properties are not yet well
fully understood. In this paper we consider a high dissipative bouncing ball model where a coefficient of restitution plays the role of dissipation, and the 
perturbation parameter is physically interpreted as a ratio between the moving plate acceleration and the gravitational field. For some 
combinations of parameters, a plenty of attractors can coexist \cite{ref31,ref32,ref33}. We found that these attractors in the phase
space are intertwined, and varying the value of the control parameter of perturbation, we characterize a boundary crisis
\cite{ref6,ref34,ref35,ref36} between the stable and unstable manifold of the same saddle point.
Such crisis leads to a successive destruction of these intertwined attractors and plays the role of a 
mechanism that allows the bottom attractor, which is related with the vibrating wall, still exist, giving to the attractor the status of a
robust one.
Also, we setup an impact physical crisis, between the real vibrating
plate and the border of an attractor. This ``unclassified impact crisis'' can work as mechanism to reduce the number of attractors, in the 
sense that before the crises we have plenty of attractors, and after it, only few of them are still present in the dynamics.

The organization of the paper is given as follows: in Sec.\ref{sec2} we describe the dynamical system under study and its chaotic
properties. Section \ref{sec3}A is devoted to the numerical analysis of the average velocity, in Sec.\ref{sec3}B we study the basin 
of attraction of the fixed points and set up the impact physical crisis, and in Sec.\ref{sec3}C we discuss the relation between the
manifolds boundary crisis and the attractors; finally in Sec.\ref{sec4} we draw some final remarks and conclusions.

\section{The model, the mapping and chaotic properties}
\label{sec2}

~~~~~In this section we describe the model under study, so called Bouncer model, which consists of a particle, under the influence
of a constant gravitational field, that suffers inelastic collisions with a heavy oscillating wall. 
Dissipation is introduced via a restitution coefficient $\gamma\in[0,1]$, where $\gamma=1$ recovers the conservative case, where
FA is inherent \cite{ref24}. The introduction of dissipation can be considered as a suppression mechanism for this unlimited energy
growth \cite{ref37,ref38}. The system is oriented along the vertical axis, where the upward direction is said to be positive, the wall
equilibrium position is set in $y=0$, and the dynamics is basically described by a non-linear mapping for the variables velocity 
of the particle $v$ and time $t$ immediately after a $n^{th}$ collision of the particle with the vibrating wall.  

There are two distinct versions of the dynamics description:
{\it(i)} complete one, which consists in considering the complete movement of the time-dependent wall, and {\it(ii)} simplified, 
where the wall is assumed to be fixed, but exchange momentum and energy with the particle upon collision.
Both approaches produce a very similar dynamics considering 
conservative \cite{ref24} and dissipative cases \cite{ref25,ref37,ref38,ref39}. In the complete version, the vibrating 
wall obeys the equation $y_w(t_n)=\varepsilon\cos{w t_n}$, where $\varepsilon$ and $w$ are respectively, the amplitude and
the frequency of oscillation of the vibrating wall. In the simplified version, the vibrating wall is said to be fixed at $y=0$, but when the
particle collides with it, they exchange momentum and energy as if the wall were vibrating. Thus, the simplified approach keeps the nonlinearity of the model and 
at the same time, significantly serves to speed up the numerical simulations, as well allows easier analytical calculations. In this paper
and from this point beyond, we only deal with the complete version of the mapping. 

Considering the flight time, which is the time that the particle spends to go up, stops with zero velocity, starts falling and
collides again with the vibrating wall, we define some dimensionless and more convenient variables as: 
$V_n=v_n w/g$, $\epsilon=\varepsilon w^2/g$, where $V_n$ is the ``new dimensionless velocity", $g$ is the gravitational field and
$\epsilon$ can be understood as a ration between accelerations of the vibrating wall and the gravitational field. For instance, one can
set some real values for the dimensional variables, as $g=10 m/s^2$, $\varepsilon=0.001m$, $w=2\pi f$, where $f=100 Hz$, and obtain the 
dimensionless variable $\epsilon\approx 0.1591$. Some real devices concerning impact experiments with granular material can be found in 
Refs.\cite{ref18,ref19}. Also, measuring the time in terms of the number of oscillations of the vibrating wall $\phi_n=w t_n$, we obtain the mapping
\begin{equation}
T:\left\{\begin{array}{ll}
V_{n+1}=-\gamma({V_n^*}-{\phi_c})-(1+\gamma)\epsilon\sin(\phi_{n+1})\\
\phi_{n+1}=[\phi_n+\Delta T_n]~~{\rm mod (2\pi)}\\
\end{array}
\right.,
\label{eq1}
\end{equation}
where the expressions for $V_n^*$ and $\Delta T_n$ depend on the kind of the considered collision.
For the case of multiple collisions inside the collision zone $[-\epsilon,+\epsilon]$, the expressions are
$V_n^*=V_n$ and $\Delta T_n=\phi_c$ where $\phi_c$ is obtained from the
condition that matches the same position for the particle and the vibrating
wall, expressed as
\begin{equation}
G(\phi_c)=\epsilon\cos(\phi_n+\phi_c)-\epsilon\cos(\phi_n)-V_n\phi_c+{{
1}\over{2}}\phi_c^2~.
\label{eq2}
\end{equation}
where this transcendental equation must be solved numerically for $G(\phi_c)=0$, with $\phi_c\in(0,2\pi]$.

If the particle leaves the collision zone case after a collision, goes up, reach null velocity, and falls for an another collision, 
we have indirect collisions and the expressions
are $V_n^*=-\sqrt{V_n^2+2\epsilon(\cos(\phi_n)-1)}$ and $\Delta
T_n=\phi_u+\phi_d+\phi_c$ with $\phi_u=V_n$ denoting the time spent by the
particle in the upward direction up to reaching the null velocity,
$\phi_d=\sqrt{V_n^2+2\epsilon(\cos(\phi_n)-1)}$ corresponds to the time that
the particle spends from the place where it had zero velocity up to the
entrance of the collision zone at $\epsilon$. Finally the term $\phi_c$ has
to be obtained numerically from the equation
\begin{equation}
F(\phi_c)=\epsilon\cos(\phi_n+\phi_u+\phi_d+\phi_c)-\epsilon-V_n^*
\phi_c+{{1}\over{2}}\phi_c^2~,
\label{eq3}
\end{equation}    
where $F(\phi_c)$ represents a transcendental equation that must be solved numerically in order to find the exact ``time'' of collision, 
as $F(\phi_c)=0$, with $\phi_c\in[0,2\pi].$

\begin{figure}[h!]
\begin{center}
\centerline{\includegraphics[width=9cm]{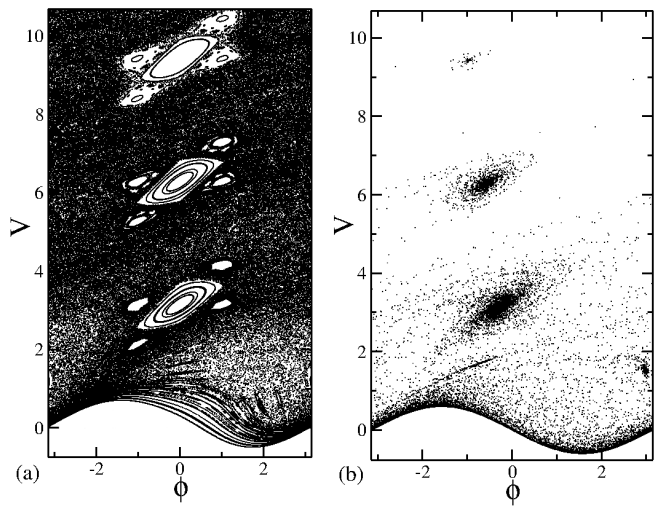}}
\end{center}
\caption{\it{Comparison between phase space for conservative and dissipative dynamics. In (a) $\epsilon=0.6$ and $\gamma=1.0$, 
and in (b) $\epsilon=0.6$ and $\gamma=0.9$. In (b) the thick black regions are the sinks and the bottom attractor. Also, all 
the spread dots are the transient.}}
\label{fig1}
\end{figure}


Taking the determinant of the Jacobian matrix of both kinds of collisions, and 
after a straightforward algebra, it is easy to show that the
mapping (\ref{eq1}) shrinks the phase space measure since the
determinant of the Jacobian matrix is given by
\begin{equation}
{\rm Det~J}=\gamma^2\left[{{V_n+\epsilon\sin(\phi_n)}\over{V_{n+1}
+\epsilon\sin(\phi_{n+1})}}\right]~.
\label{eq5}
\end{equation}
Here, if $\gamma=1$ we recover the non-dissipative version of the mapping, in fact, as velocity and phase are not canonical pairs
in the complete version, the determinant of $J$ is not the unity, but rather it leads to the following measure to be preserved, 
$d\mu=(V+\epsilon\sin\phi)dV d\phi~$. Indeed, the extended phase space for the 
whole version of the model considers four variables namely: (1) $y_w$ denoting the position of the vibrating wall; (2)
$V_p$ corresponding to the velocity of the particle; (3) $E_p$ which is the mechanical energy (kinetic+gravitational)
of the particle and (4) the time $t$. The canonical pairs however are: position and velocity $(y_w,V_p)$ and; energy and time $(E_p,t)$.

Another useful property for the dynamics evolution, as function of the control parameters, is the analysis of the fixed points and
their stability. For the Bouncer model the period-1 fixed points can be obtained by doing 
$V_{n+1}=V_n=V^*$ and $\phi_{n+1}=\phi_n=\phi^*+2m\pi$ in the Eq.(\ref{eq1}). For both kinds of collisions, successive and indirect, 
the fixed points are 
\begin{equation}
V^*=m\pi~; m=1,2,...~~\phi^*=\arcsin \left({V^*(\gamma-1) \over (1+\gamma)\epsilon} \right)~.
\label{eq6}
\end{equation}

Their stability are given by $Det(J-\lambda I)=0$, evaluated over the fixed points, where $\lambda$ are the eigenvalues
of the Jacobian matrix and $I$ is the identity matrix. The eigenvalues can be found solving the expression
$\lambda_{1,2}={TrJ\pm\sqrt{{TrJ}^2-4DetJ}\over2}$ \cite{ref2}.\\ 

Figure \ref{fig1} shows a comparison between the phase space for the conservative and dissipative versions. As the 
dynamics evolves, the introduction of dissipation destroys the invariant curves in the stability islands,
and the stable fixed points become sinks \cite{ref1,ref2,ref6}. Depending of the control parameter, there may have a
plenty of these attractors, where the orbits converge to them. In Fig.\ref{fig1} one can see that after the dissipation was 
introduced the first three stability islands, denoted by white regions among the chaotic sea in Fig.\ref{fig1}(a), became attracting 
fixed points. Also, there is the presence of an attractor on the bottom of Fig.\ref{fig1}(b), near where the vibrating wall is located.
For a better understanding and visualization, we are going to use in all figures the phase representation between $-\pi$ and $+\pi$.\ 

\section{Results and Discussions}
\label{sec3}

In this section, we describe the results obtained by the numerical simulations. First we draw some average velocity 
curves for a combination of the two control parameters, $\epsilon$ and $\gamma$. Then, the basins of attraction of some
attracting fixed points are obtained. We investigate how these basins of attraction behave, as we range the control
parameters. We constructed some bifurcation diagrams for the fixed points and analysed how their stability vary. By
drawing the unstable and stable manifolds we notice that the attractors are intertwined in the phase space, and there is
a boundary crises, a crossing of their manifolds, which creates a successive mechanism of destruction for all attractors originated
by the sinks. Also, we made a study over the bifurcation process and the chaotic properties of it.

\subsection{Average Velocities}

Let us set the equation for the average velocity, which depends on both $\epsilon$ and $\gamma$. The statistics was made considering
two steps, an average taken along the orbit, evolved until a finite high number of collisions $n$ and 
an average taken along the ensemble of initial conditions. So, we may define

\begin{figure}[h!]
\begin{center}
\centerline{\includegraphics[width=9.5cm,height=10.0cm]{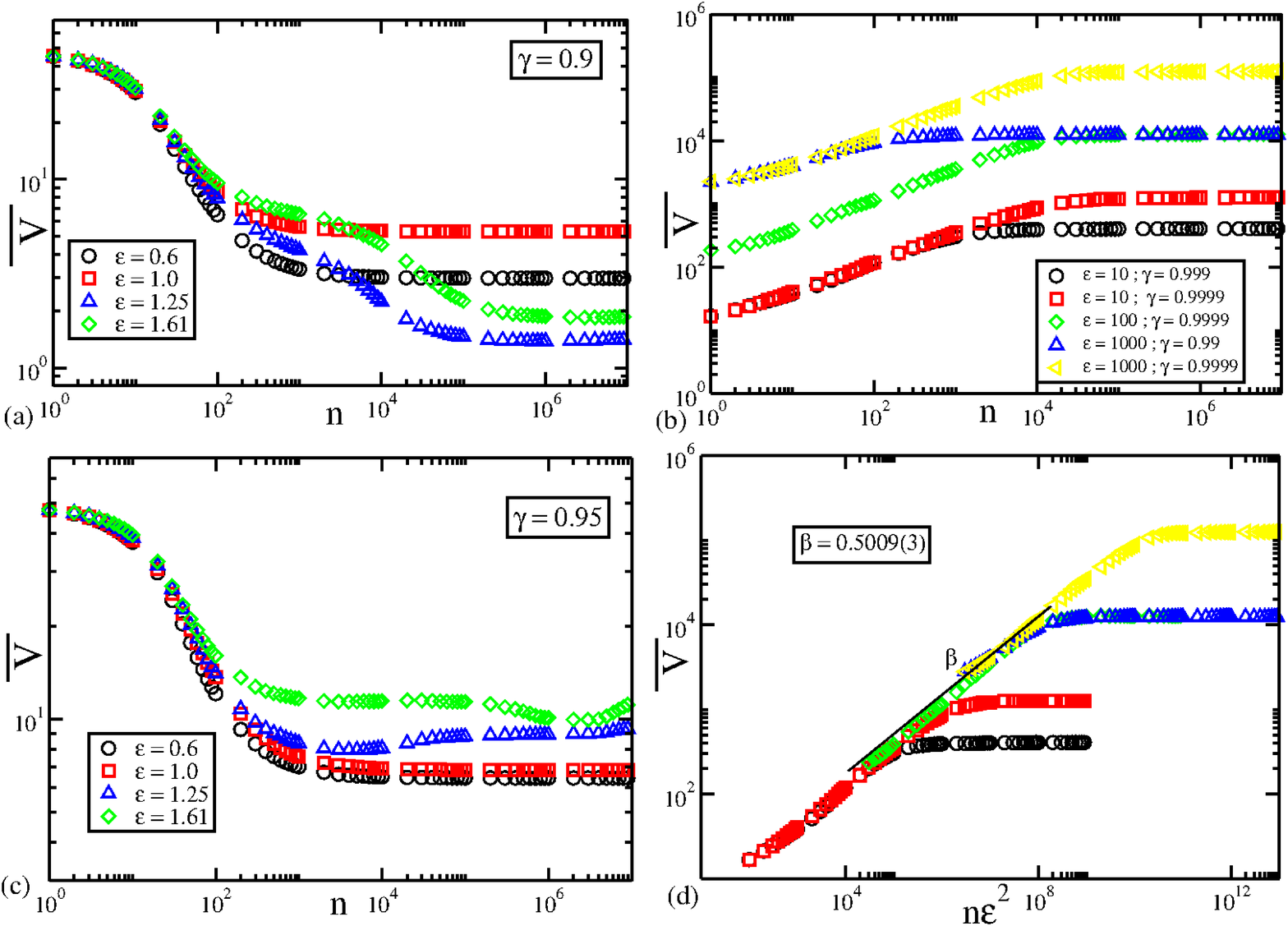}}
\end{center}
\caption{Color online: \it{Behaviour of the average velocity curves for an extensive range of $\epsilon$ and $\gamma$. In (a) and (c), we have
a high dissipation and low $\epsilon$, and the dynamics are basically controlled by the attracting fixed points. In (b) and (d), 
we have small dissipation and high $\epsilon$. Here the dynamics is ruled by the attractor localized in the bottom, 
embedded with the vibrating wall. Also, in (d) a rearrange in the horizontal axis is made, in order to make all the curves
start to grow together. This rearrange is convenient, when we make use of the scaling techniques.}}
\label{fig2}
\end{figure}

\begin{equation}
V_i(n,\epsilon,\gamma)={1\over n} {\sum_{j=1}^n} V_{j}~,
\label{eq8} 
\end{equation}
and hence
\begin{equation}
\overline{V}={1\over M} {\sum_{i=1}^M} V_i(n,\epsilon,\gamma)~,
\label{eq9} 
\end{equation}
where $M$ represents an ensemble of initial conditions. The index $j$ and $i$ represent respectively the collision number, and the number of 
initial conditions.\ 

Figure \ref{fig2}, shows the evolution of the average velocity for an extensive range of the control parameters $\epsilon$ and
$\gamma$ as function of the number of collisions. Here, we consider two regimes: {\it(i)} high dissipation and small $\epsilon$,
and {\it(ii)} small dissipation and high $\epsilon$. One can identify very distinct behaviour between both regimes. Considering 
regime {\it(i)}, as shown in Figs.\ref{fig2}(a,c), we see that the $\overline{V}$ curves start with high initial velocity 
near $V_0$, given as an initial condition, and experience an exponential decay, \cite{ref39}. After a transient
they bend towards different regimes of saturation in low energy levels. Basically, the orbits are attracted by the several fixed
points that coexist in the phase space, when the perturbation parameter $\epsilon$ is still small enough, therefore marking
their convergence to different plateaus. On the other hand, in regime {\it(ii)}, as present in Fig.\ref{fig2}(b,d), the 
$\overline{V}$ curves start with low initial velocity, near $V_0$ given as an initial condition, and they experience a growth
for short time according to a power law with exponent $\beta\approx0.5$, until they bend towards a saturation regime, in a 
high energy levels. For this case, the saturation plateaus of high energy are connected with the chaotic attractor,
and there are no attracting fixed points for such high energy regime. Also, if we rescale the
horizontal axis by $n\epsilon^2$ all $\overline{V}$ curves, start to grow together. For a more complete investigation on 
the {\it(ii)} regime concerning the chaotic attractor behaviour and a fully scaling analysis on the $\overline{V}$ curves,
we recommend Refs.\cite{ref37,ref38} for a numerical point of view, and Refs.\cite{ref25,ref39} for an analytical interpretation.
However, in this paper, we are interested in the regime of high dissipation
and small $\epsilon$, shown in Figs.\ref{fig2}(a,c), where the attracting fixed points still exist and play an important role 
in the dynamics. Indeed, this high dissipation analysis can be very useful in a further experimental analysis of the Bouncer model, 
once they are easily obtained, instead of the tiny dissipations as used in Refs.\cite{ref25,ref37,ref38,ref39}, that would be 
impossible to obtain in a laboratory. Also, one could think about experience with granular materials interacting with 
vibrating plates \cite{ref15,ref16,ref17,ref18,ref19}, as a direct application of the Bouncer model and its phenomena.

\subsection{Basins of Attraction and Bifurcations}

\begin{figure}[h!]
\begin{center}
\centerline{\includegraphics[width=9cm]{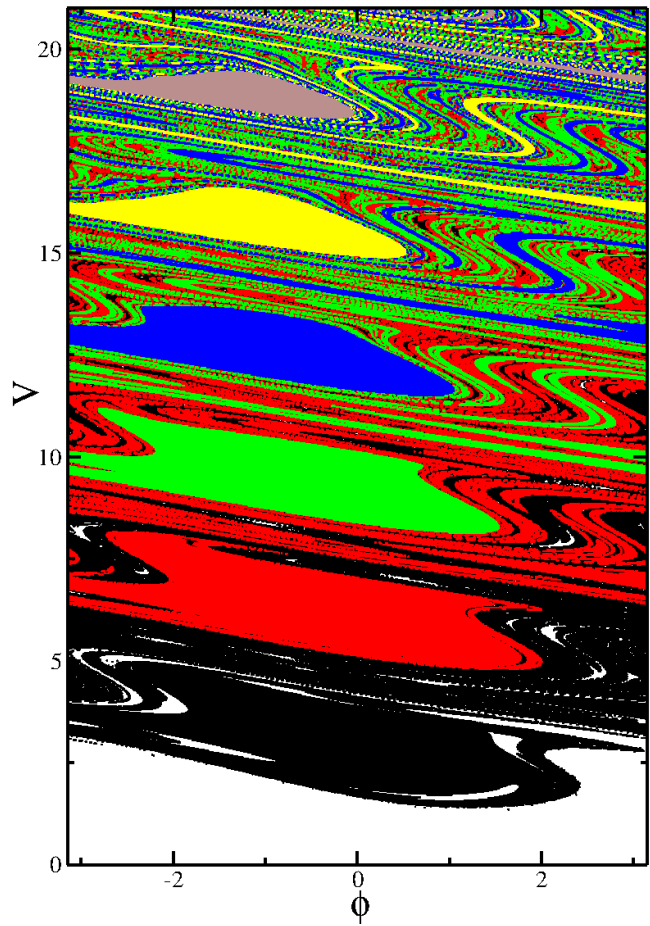}}
\end{center}
\caption{Color online: \it{Basin of attraction for the fixed points considering $\epsilon=1.0$ and $\gamma=0.95$. The colors represent
the fixed points were the initial conditions are being attracted to. Black $\rightarrow \pi$, red $\rightarrow 2\pi$, green, $\rightarrow 3\pi$, blue $\rightarrow 4\pi$, 
yellow $\rightarrow 5\pi$ and brown $\rightarrow 6\pi$, and the white regions denotes initial conditions that converged to the bottom attractor.}}
\label{fig3}
\end{figure}

\begin{figure*}
\begin{center}
\centerline{\includegraphics[width=18cm,height=13.0cm]{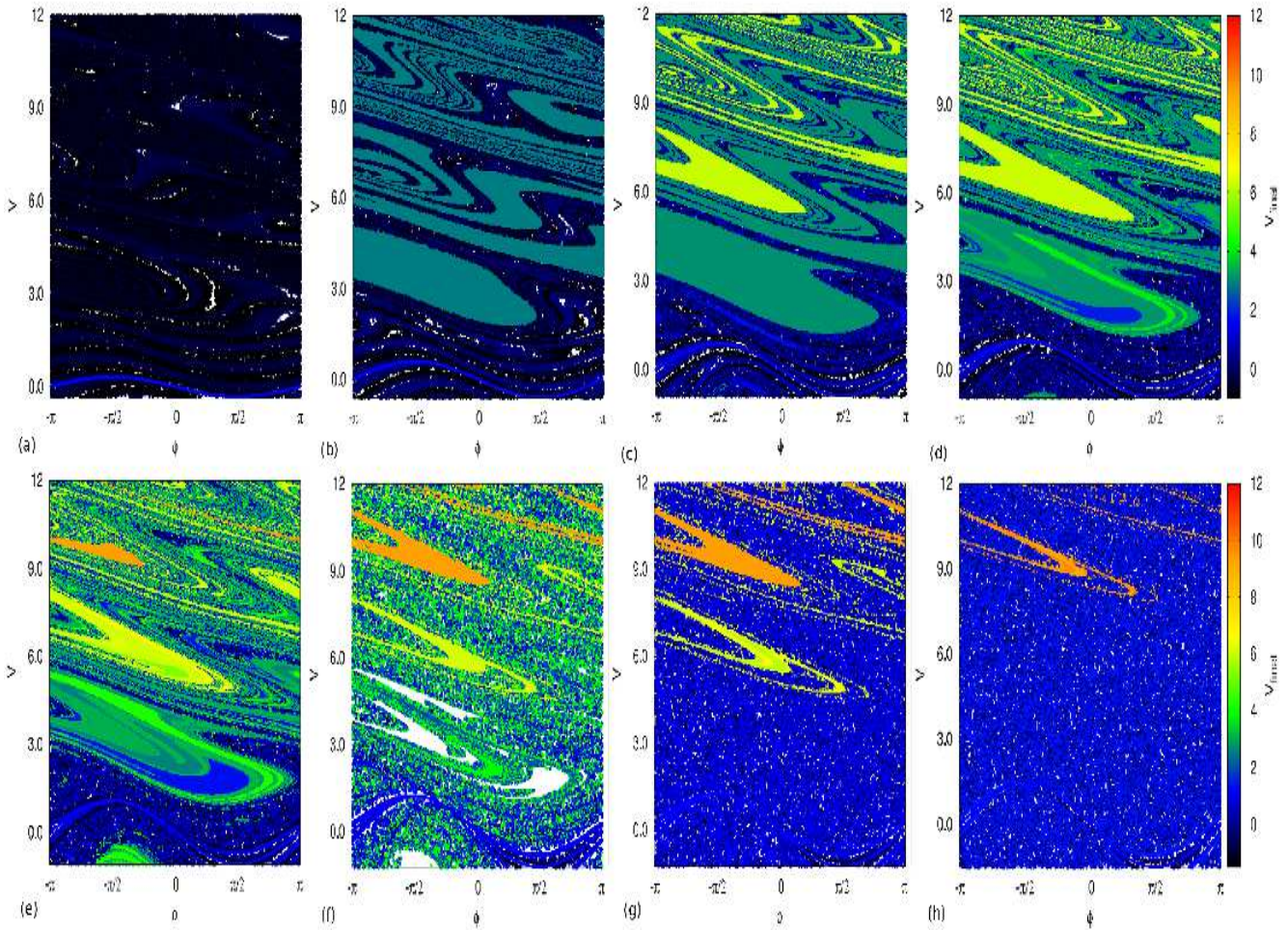}}
\end{center}
\caption{Color online: {\it Evolution of the basin of attraction for $\gamma=0.8$ for a grid of $500\times500$ initial conditions.
The perturbation parameter follows as: (a) $\epsilon=0.32005$, (b) $\epsilon=0.60095$, (c) $\epsilon=0.88715$, (d) 
$\epsilon\approx\epsilon_t=0.964$,
(e) $\epsilon=1.08590$, (f) $\epsilon\approx\epsilon_c=1.22635$, (g) $\epsilon\approx\epsilon_d=1.25815$ and (h) $\epsilon=1.50195$.
The color scale denotes the final 
velocity of each pair of initial condition after a transient of $10^5$ iterations. One can see that the basins of attraction 
suffer huge transformations, as successively bifurcation process as $\epsilon$ is increased, until they are destroyed when 
$\epsilon$ acquires critical values.}}
\label{fig4}  
\end{figure*}

Depending on the combination of the control parameters $\epsilon$ and $\gamma$, several attracting fixed points can coexist 
in the phase space, where some of them can be more influential to the dynamics than other. In order to understand how these attracting
fixed points behave as related to initial conditions, Fig.\ref{fig3} shows a basin for the periodic attractors for $\epsilon=1.0$ and 
$\gamma=0.95$, where a grid of $1000\times1000$ initial conditions were equally split and set in the axis of velocity 
$V\in[-\epsilon,7\pi]$
and phase $\phi\in[-\pi,+\pi]$. The different colors represent the basins of attraction for each attractor of period-1 located in $V^*=m\pi$,
according the fixed points obtained in Eq.(\ref{eq6}). Here, each initial condition were evolved up to a transient of $10^5$ collisions, 
and then we marked its final velocity in the phase space. 

Since, we have a positive restitution coefficient, one may think that the 
particle would ``glue" on the vibrating wall for long times. This indeed can happen, if it lands deep enough in the absorbing region of the 
phase space, the particle will perform a large number of smaller and smaller bounces, that could follow progressive geometric conversions 
\cite{ref30}. Such peculiar behaviour is known as locking \cite{ref30}. The white regions in Fig.\ref{fig3} denotes the initial
conditions that converged to the locking region attractor, i. e. the bottom attractor embedded with the vibrating wall. However, if the
particle has a positive relative velocity, it will not be glued to the wall. And, depending of the on the control
parameters, this velocity can acquire multiplicity, in different regions of the phase space, giving birth to periodic attractors.
In Fig.\ref{fig3}, each color denotes a different period-1 fixed point basin of attraction. As $m$ is increased, it seems that the basins
are getting smaller, which could be a possible indication of less influence in the dynamics. Also, for the higher values of $m$, the boundaries of
the basins, that are limited by the unstable manifolds of each fixed point \cite{ref1,ref6,ref8}, behave
in a very complicated stretching and mixing way, folding themselves like a horseshoe \cite{ref1,ref6,ref8}.

The beautiful mazy behaviour of the basins of attraction shown in Fig.\ref{fig3}, can drastically change, when $\epsilon$ is increased.
As one can see in Fig.\ref{fig4}, where the evolution of the basin of attractions for $\gamma=0.8$
and some values of $\epsilon$ is shown. Each item of Fig.\ref{fig4} was constructed considering
a grid of initial conditions of $500\times500$ equally distributed along the velocity $V\in[-\epsilon,3\pi]$ and phase 
$\phi\in[-\pi,+\pi]$. The color scale represents the final velocity of each pair of initial condition after a transient of $10^5$ 
iterations. The color scale was kept fixed as the final velocity ranges. For low velocities and orbits that were attracted 
to the bottom attractor, we have the darker colors as black, dark blue (black). The orbits that were attracted by the 
first two sinks, $V^*=\pi$ and $V^*=2\pi$, the color scale ranges from blue (dark gray) to yellow (light gray), and for the high velocity 
attracting orbits (said above $V^*=2\pi$), we have the color range scale between yellow (light gray) and red (dark gray).\

\begin{figure}[h!]
\begin{center}
\centerline{\includegraphics[width=9cm,height=12cm]{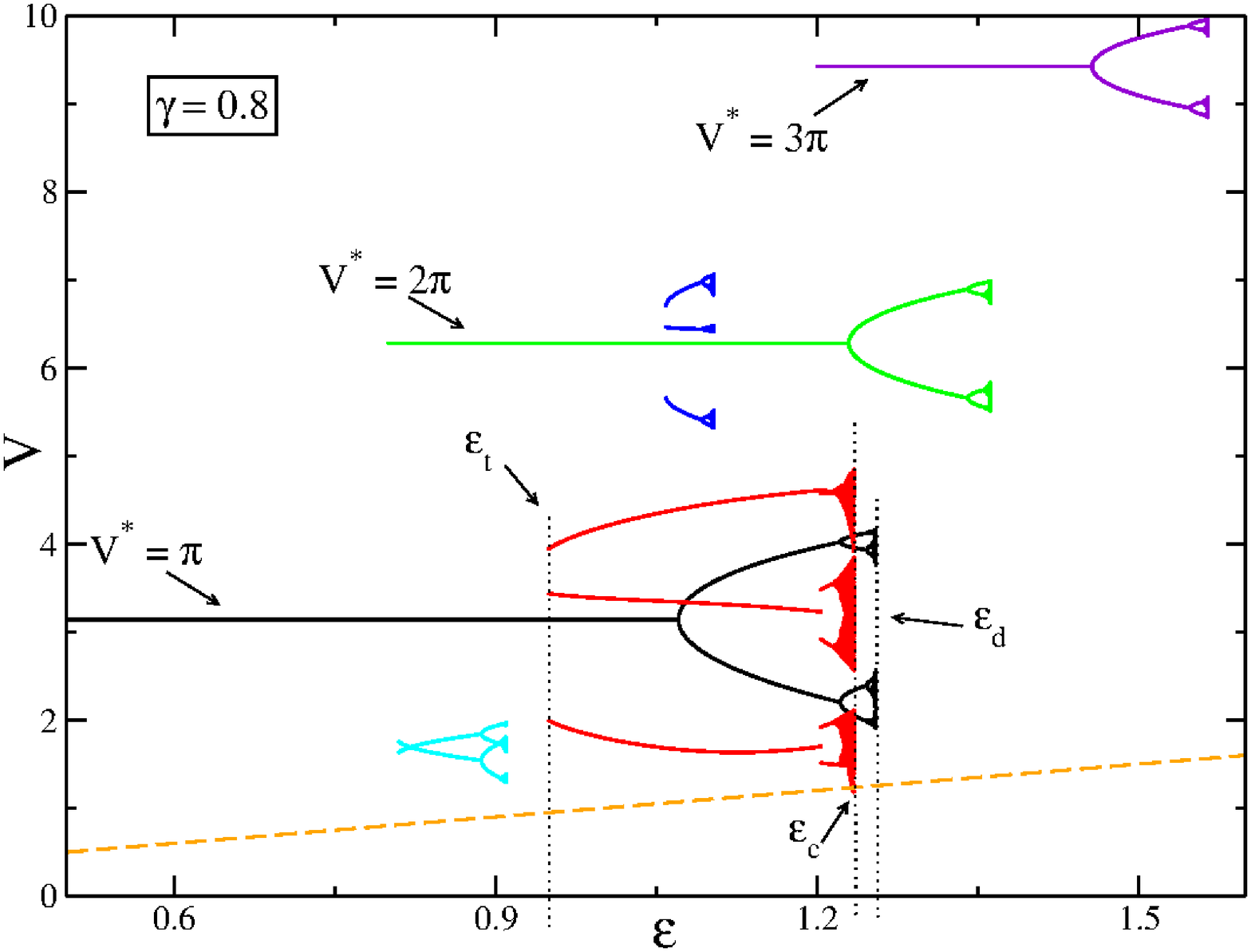}}
\end{center}
\caption{Color online: \it{Bifurcation diagram as function of $\epsilon$ for $\gamma=0.8$. It is shown a scenario of the 
evolution of the attracting fixed points, where attractors originated by the sinks $V^*=m\pi$, coexist with period-3 orbits originated by
tangent bifurcations, the bottom attractor (given by the dashed line) and others attractors. This figure illustrates the plenty of attractors
and the high active dynamic structure that orbits may experience. Also, three critical values of $\epsilon$ concerning the crises are 
characterized.}}
\label{fig5}
\end{figure}

In Figure \ref{fig4}, as $\epsilon$ is increased, the boundaries of the basins of attraction start to grow
following the stretching and mixing behaviour, just like the Smale horseshoes, \cite{ref1,ref6,ref8}, as
Figs.\ref{fig4}(a,b,c) displays, where in particular for Fig.\ref{fig4}(a), the sink located in $V^*=\pi$ did not exist yet, the same
applies for Fig.\ref{fig4}(b), where the sink located in $V^*=\pi$ also did not exist. One can check that by the period-1 one fixed points
expressions, specially for $\phi^*$. As we increase $\epsilon$, the sinks 
located in $V^*=m\pi$ start to notice the consequences of being lied intertwined with the bottom attractor. Considering 
the first sink $V^*=\pi$, it suffers a tangent bifurcation in $\epsilon_t\approx0.964$ (this bifurcation will be explained in the manifolds section),
which creates three new attracting zones inside its own basin of attraction, as shows Fig.\ref{fig4}(d). Raising the value of $\epsilon$ 
a little bit, the sink in $V^*=\pi$, bifurcates, creating, together with the other sinks, a plenty of attracting regions in the phase space,
as shown in Fig.\ref{fig4}(e). The crises happens near $\epsilon_c\approx1.22635$, shown in Fig.\ref{fig4}(f). After the tangent bifurcation in 
$\epsilon_t$, three branches evolves as $\epsilon$ increases. The two upper branches collide with each other, generating a boundary 
crises that destroys both branches. At the same parameter $\epsilon_c$, the lower branch, suffers a physical collision with the
vibrating boundary. This is a different crises, once the attractor is colliding with a physical structure, instead of a another attractor
or manifold. This non categorized crises can be better visualized in Fig.\ref{fig5}. Finally, in Figs.\ref{fig4}(g,h), the basin of 
attraction of the fixed point $V^*=\pi$ (said together with the branches of the tangent bifurcation) is totally destroyed, and the 
other basins of attraction for the other upper sinks started their successive destruction process. In the end, when we have a large 
enough value of $\epsilon$, only the bottom attractor remains in the system.\
\begin{figure*}
\begin{center}
\centerline{\includegraphics[width=17cm,height=12.0cm]{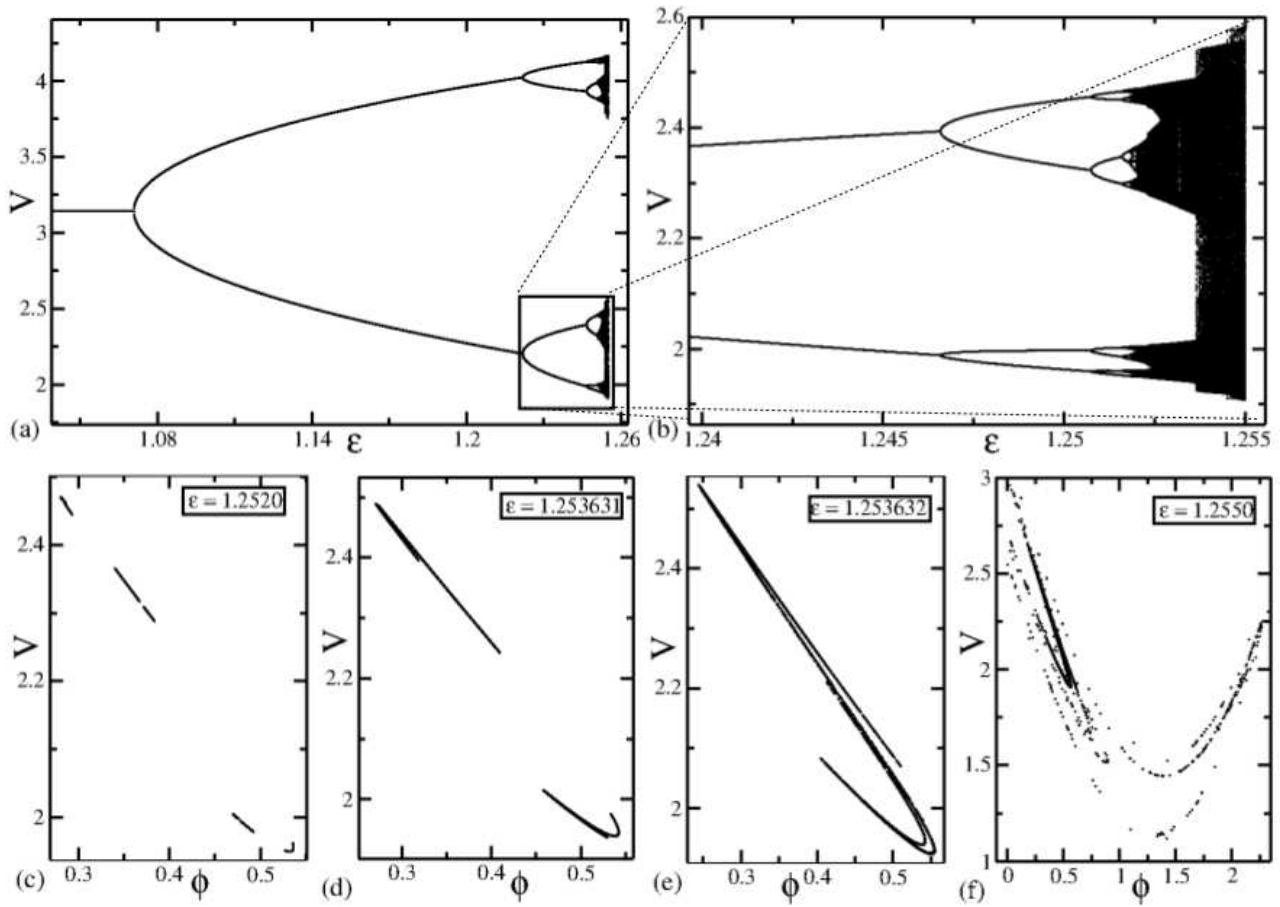}}
\end{center}
\caption{{\it In (a) we show an amplification of the successive bifurcation process that the first sink suffers as $\epsilon$ is increased, 
in (b) we made a bigger zoom yet in this process. In (c), (d), (e) and (f), we display the shape of the local chaotic attractors in 
the phase space, for the lower branch of the first sink. The values of $\epsilon$ used were: 
in (c) $\epsilon=1.2520$, (d) $\epsilon=1.253631$, (e) $\epsilon=1.253632$,  and in (f) $\epsilon=1.2550$. 
One can see that the attractors pass through a bifurcation process, and after a collision between the branches, they merged into
a bigger chaotic attractor. In all items we considered $\gamma=0.8$.}}
\label{fig6}  
\end{figure*}

For a better understanding and visualization of the crises, sinks bifurcations and the evolution of the basins of attraction, we constructed
a bifurcation diagram for some fixed points, as shows Fig.\ref{fig5}, for a fixed dissipation $\gamma=0.8$. The diagram illustrates the 
plenty of attractors and the high active dynamic structure that orbits may experience during the time evolution. This diagram was constructed in 
two different ways: {\it (i)} we have an initial condition very near the sinks, and let it follows the attractor, as $\epsilon$ increases;
and {(\it ii)} we kept the same initial condition for every value of $\epsilon$. In both cases, the range of $\epsilon$ was split in $5000$ 
equal parts. Considering the evolution in {(\it i)}, the dynamics follows a regular bifurcation diagram, where the period-1 sinks located in
$V^*=m\pi$, stay stable until them suffer successive bifurcations as $\epsilon$ increases, until it finally disappear.
In the same manner, tangent bifurcations happen for $V^*=\pi$, $V^*=2\pi$ and so on. These new branches for each tangent bifurcation, 
have the same fate of the attractors originated by the sinks. These fixed points of the tangent bifurcation were obtained considering case
{\it(ii)}, where the evolution of the same initial condition, gives rise to different attractors in the bifurcation diagram, for example,
the one located near $V^*\cong1.7$, basically suffers the same bifurcation process as the main ones and then it is destroyed. 

One can notice in Fig.\ref{fig5}, that there are three critical values: {\it(i)} $\epsilon_t\approx0.964$, which is the value of $\epsilon$ where
the tangent bifurcation occurs for the first sink $V^*=\pi$; {\it(ii)} $\epsilon_c\approx1.22635$, which is the critical parameter where the crises
occurs for the three branches originated in the tangent bifurcation. One can see in Fig.\ref{fig5}, that in $\epsilon=\epsilon_c$, there are
two simultaneous crisis. The upper two branches collide, and destroy each other, and at the same "time", the lower branch, physically collides
with the vibrating wall, denoted here by the dashed line, characterizing an yet unclassified physical crises, between the real structure 
(vibrating wall), and the border of an attractor. Finally, {\it(iii)} $\epsilon_d$,
is the value where the attractor originated by the sink $V^*=\pi$ is destroyed. Beyond $\epsilon_d$, a successive destruction mechanism takes 
place in the dynamics, destroying all the other attractors, except by the bottom one, that starts to rule the dynamics. The values of 
these critical $\epsilon$ parameters may vary as we range the dissipation.

\begin{figure*}
\begin{center}
\centerline{\includegraphics[width=17cm,height=7.0cm]{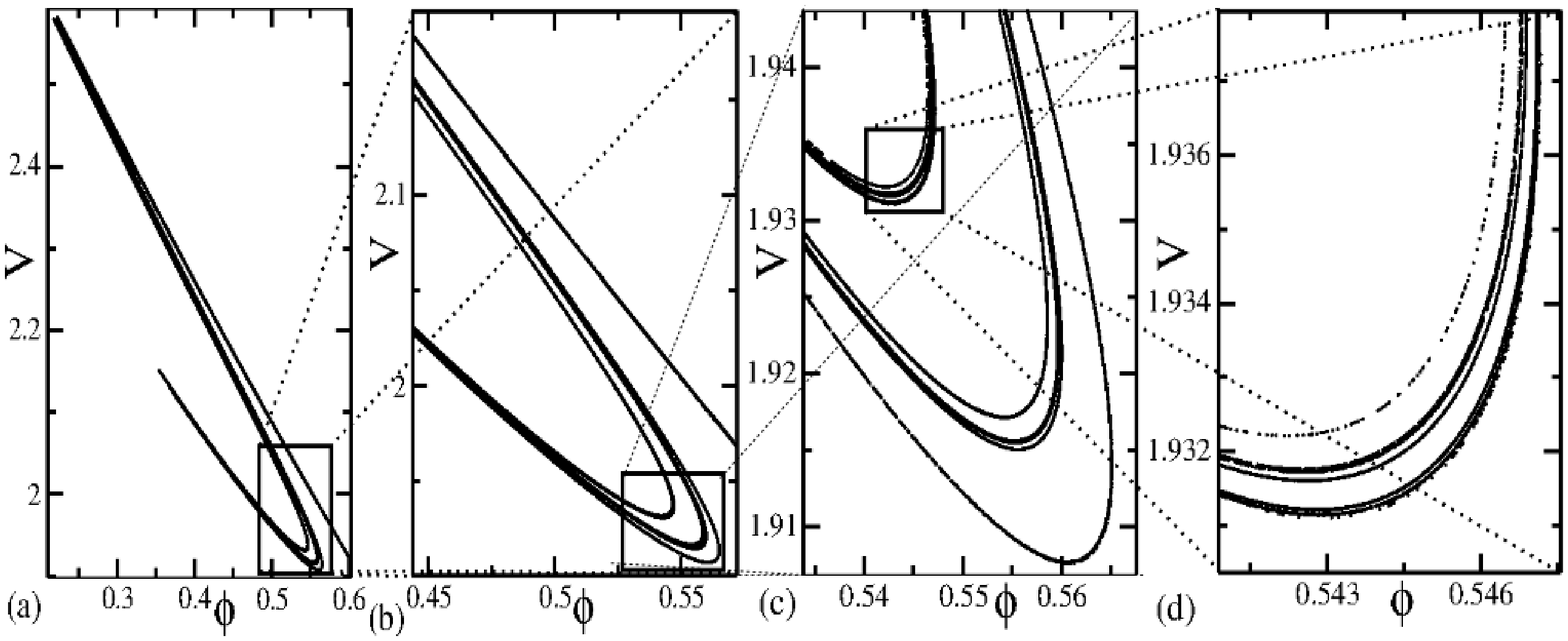}}
\end{center}
\caption{{\it Successive zoom-in windows of the chaotic attractor for $\epsilon=1.2549$. It seems that the chaotic attractor has a fractal-like 
shape very similar to one found in the Henon-Heiles map \cite{ref30}.}}
\label{fig7}  
\end{figure*}

Let us now address to the chaotic attractor born from the bifurcation process of the fixed point $V^*=\pi$. As one can see from
Fig.\ref{fig5}, when the bifurcation process occurs, we have the creation of two symmetric branches originated in the main
bifurcation. According to Fig.\ref{fig5}, each branch, will begin its own bifurcation process, which further will collide 
with itself in a boundary crises, generating its own local chaotic attractor. Figure \ref{fig6} shows how these processes occur 
for the first sink, presenting the shape of the attractors in the phase space, also we compare it with amplifications of the 
bifurcation diagram. Once, both branches are symmetric, let us do this analysis considering just one of them.
One can see in Figs.\ref{fig6}(a,b) the behaviour of bifurcation process of the branches as $\epsilon$ is increased.
Comparing them with Fig.\ref{fig6}(c), for $\epsilon=1.2520$, we can see that the attractor is under a period-8 bifurcation.
Increasing $\epsilon$, and after a comparison between Figs.\ref{fig6}(a,b) and Figs.\ref{fig6}(d,e); the two branches
of the attractor merge together into one local chaotic attractor. Here we can characterize another crisis between the attractors,
where two attractors become one. Such crisis is known as fusion one \cite{ref1,ref6}. Finally, in Fig.\ref{fig6}(f) we 
have the behaviour of the local chaotic attractor in its final stage for $\epsilon=\epsilon_d\approx1.2550$, where it seems to take the
shape of the old basin of attraction, originated by the tangent bifurcation, and then it vanishes.
Here, another crises can be characterized. The attractor (darker region), collides with the transient basin of the old attractor already
destroyed (originated by the tangent bifurcation). This ghost behaviour of the transient basin \cite{ref6}, allows to the attractor
to interact with the region that suffered the physical collision with the vibrating wall. This interaction, destroys the attractor 
in the same manner that destroy the previous one in $\epsilon\approx\epsilon_c$.

One can ask about the nature of this local chaotic attractor. Indeed, it seems to have fractal shape, as one can see in
the successive amplification of Figs.\ref{fig7}(a,b,c,d). The shape of the chaotic attractor is basically the same, no matter how 
much further in the zoom windows. It is interesting, the fact that the kind of crises we are seeing here, are basically the
same ones found in the Henon-Heiles attractor \cite{ref6,ref40,ref41}, where the chaotic attractor collides with the
stable manifold and is destroyed. Also, the same fractal-like shape of the chaotic attractor is found. It would be interesting to 
investigate later, if any other chaotic property comom related between both attractors can be found.


\subsection{Manifolds}
Let us address now to the crises related with the manifolds. It is well known in the literature that a saddle fixed point,
in the plane $V~vs.~t$ has stable and unstable manifolds \cite{ref1,ref6,ref7,ref8,ref28,ref34,ref35,ref36}.
The stable manifolds are formed by a family of trajectories that turn away from the saddle fixed point. 
One of them evolves to the chaotic bottom attractor, or visit the region of the 
chaotic bottom attractor after the event of crisis; while the other one evolves towards an attracting fixed point.
These unstable manifolds are obtained from the iteration of the map $T$ with appropriate initial conditions. 
Similarly, the construction of stable manifolds are a little bit more complicated since the inverse of the mapping, 
say $T^{-1}$ , must be obtained. Here the operator follows $T^{-1}(V_{n+1},\phi_{n+1})=(V_n,\phi_n)$. So after some  algebra, we have

\begin{figure}[h!]
\begin{center}
\centerline{\includegraphics[width=8.5cm,height=12cm]{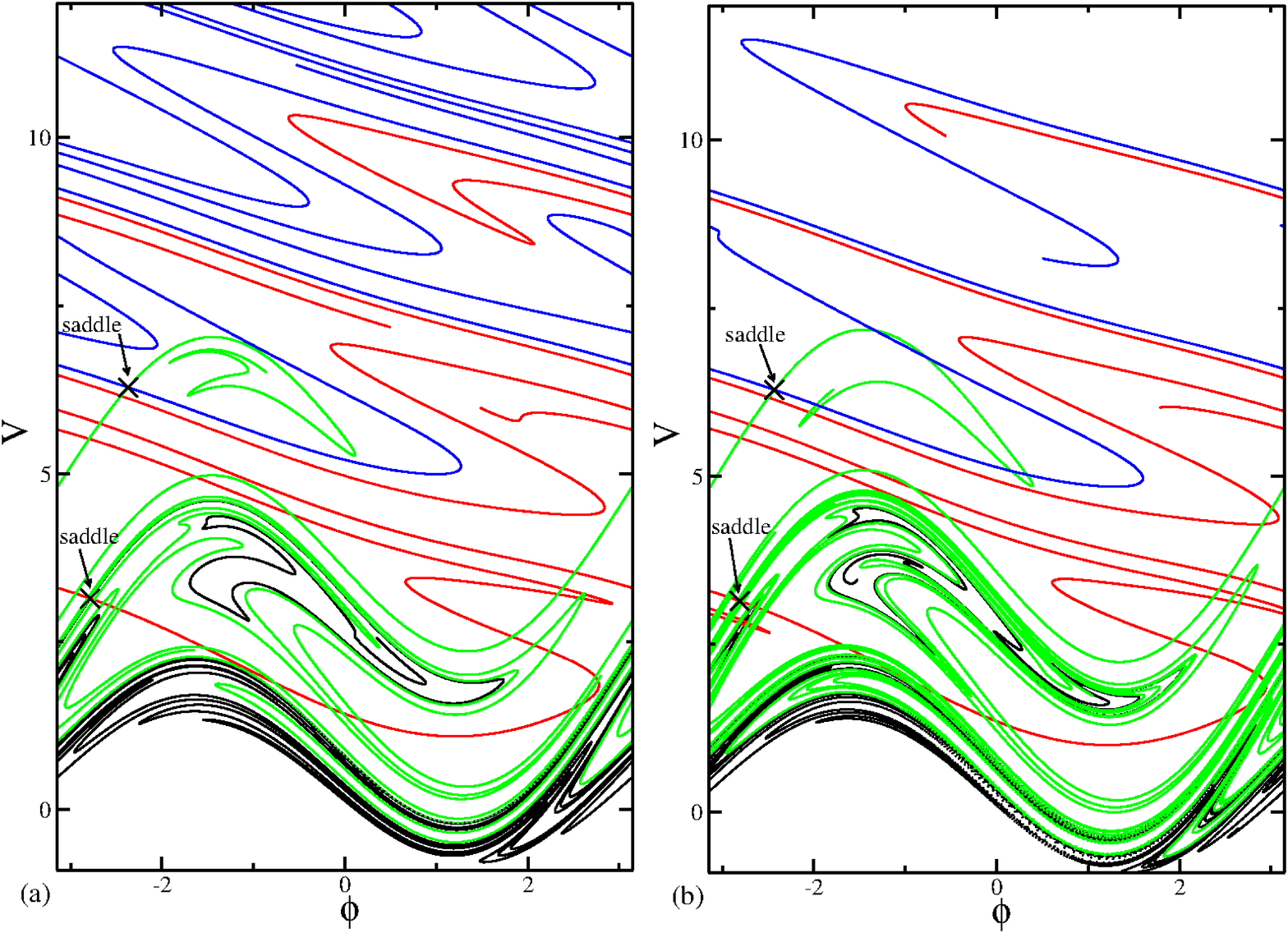}}
\end{center}
\caption{Color online: \it{Boundary crises and the embedded behaviour of the stable and unstable manifolds for $\gamma=0.8$. In (a) $\epsilon=1.0$ 
and in (b) $\epsilon=1.07$. One can see that in (a) there is no crossing between
the manifolds of either saddle fixed points, and in (b) the boundary crisis occurred for both saddle points.}}
\label{fig8}
\end{figure}

\begin{equation}
T^{-1}:\left\{\begin{array}{ll}
\phi_n=\phi_{n+1}-V_n-\nu~\\
h(V_n)=V_{n}^2+2\epsilon[\cos(\phi_{n+1}-V_n-\nu)~\\
-\cos(\phi_{n+1})]-\nu^2~\\
\end{array}
\right.,
\label{eq7}
\end{equation}
where $\nu=[V_{n+1}+(1+\gamma)\epsilon\sin(\phi_{n+1})]/\gamma$. Here, the function $h(V_n)$ must be solved numerically, 
once it depends on both $V_n$ and $V_{n+1}$. So, we made use of
the Newton's method to find the root, where $V_n=V_{n+1}-h(V_n)/h^{\prime}(V_n)$, and 
$h^{\prime}(V_n)=2V_n+2\epsilon\sin(\phi_{n+1}-V_n-\nu)$.\

The procedure for obtaining the stable manifolds is the same as
that one used for the unstable manifolds, however, instead of iterating the map $T$ we must iterate its inverse $T^{-1}$. We set
a tiny circle of radius $\delta=10^{-4}$, and split $10^4$ initial conditions around the respective saddle point and iterate the normal and reverse dynamics. After that, 
we make a zoom-in the region of the saddle point, and made a linear fit in both branches of the unstable and stable manifolds, in
order to find the eigenvectors. After that, we just evolve the normal and reverse dynamics again, but considering the distribution
of initial conditions along the linear fit of the manifolds branches. Just for notice the saddles are localized in $V^*=m\pi$ 
and $\phi^*\rightarrow\phi^*-\pi$.\

For closed domain dynamical systems, such as the Fermi-Ulam model and other time-dependent billiards \cite{ref42,ref43,ref44,ref45}, 
the unstable manifolds generate the border of the  basin of attraction of the chaotic attractor and the stable manifolds draw
the boundaries of the basin of the attracting sinks, a boundary crises happens when the stable manifold touches the 
unstable manifold of the same saddle fixed points due to a modification of the control parameter. In this case, the chaotic 
attractor is destroyed, and only the sink remains in the system. This collision implies in a sudden destruction of the chaotic 
attractor and also of its basin of attraction \cite{ref34,ref35}. This destruction can be very useful as a mechanism of
controlling chaos in this dissipative version of the model since, after the crisis event, the particle is captured by an 
attracting fixed point (sink). However, the kind of crisis that happens in the unbounded Bouncer model is a bit more complicated,
once we have plenty of attractors, and their manifolds found themselves embedded.

Stable and unstable manifolds for the first two saddle points for a dissipation value of $\gamma=0.8$ are displayed in Fig.\ref{fig8}(a,b).
Both branches of the stable manifold behave as follows: the upward branch evolves to the attracting fixed point while the downward 
branch evolves to the chaotic attractor. In both Figs.\ref{fig8}(a,b), we see that the stable manifold for the saddle in $V^*=2\pi$, 
are embedded with the stable manifold of the saddle $V^*=\pi$. Both stable manifolds are also intertwined with the attractor in the bottom.
We believe that this peculiar behaviour happens for all the saddles located above in the phase space for all values of $V^*=m\pi$,
creating a whole chain of iteration between the attracting fixed points and the attractor in the bottom.
Also, in both Figs.\ref{fig8}(a,b) the two branches of the unstable manifold generate the basin boundaries for both attracting fixed points.
Here, there is no iteration between the unstable manifolds of different sinks. The unstable manifolds go up and up in the velocity axis, in 
a stretching and mixing way, drawing the limits of the attracting boundaries of each fixed point. One can imagine how they would 
behave just looking at Fig.\ref{fig3}, where the basin of attraction until $V^*=6\pi$ is drawn. It would be interesting to compare 
the relations between the stretching and mixing of the manifolds with the Smale horseshoe mapping.
 
Also, in Fig.\ref{fig8}(a) shows the embedded manifolds $\epsilon=1.0$. One can notice here, that there is no crossings between the boundaries
of the stable and unstable manifold of both saddle points yet, but we can visualize the tangent bifurcation occurring in the sink located in
$V^*=2\pi$, where three branches rise and start to grow from the place where the sink should be located. One can also compare with 
Fig.\ref{fig5}, and confirm that this is the exactly control parameter of the tangent bifurcation.
Once we raise the value of the control parameter $\epsilon$, the respective unstable and stable manifold from both saddle points cross
each other, as show Fig.\ref{fig8}(b), for $\epsilon=1.07$. One could imagine that the crossing behaviour of the manifolds would destroy
the attractor in the bottom, and let only the sink in the phase space. This indeed happens, but once the manifolds for all saddles found
themselves embedded, this destruction seems not to cause greater effects in the dynamics. Because even that for one saddle the attractor is
destroyed, there will be the manifold from upper saddle points, embedding themselves with the region where the attractor was, giving an
"extra life time" for the attractor, until the next boundary crises from the upper saddles, where the same process will occur in a successive way.
We think, that the boundary crises between the manifolds, are serving as a trigger for the unbounded
growth of the chaotic attractor in the bottom. One could imagine for higher values of $\epsilon$, that the bottom chaotic can be very 
influential in the dynamics. If, we take $\epsilon=10$, the bottom chaotic attractor would occupy the region where the first three sinks 
of period-1 should be located, and the crossing between the manifolds that we are seeing here in Fig.\ref{fig8} for low values of $\epsilon$,
would be happening for very high saddle points in the phase space. We dare to call this mechanism, as a regenerating process, or 
``robust attractor".

\begin{figure}[h!]
\begin{center}
\centerline{\includegraphics[width=8.5cm,height=12cm]{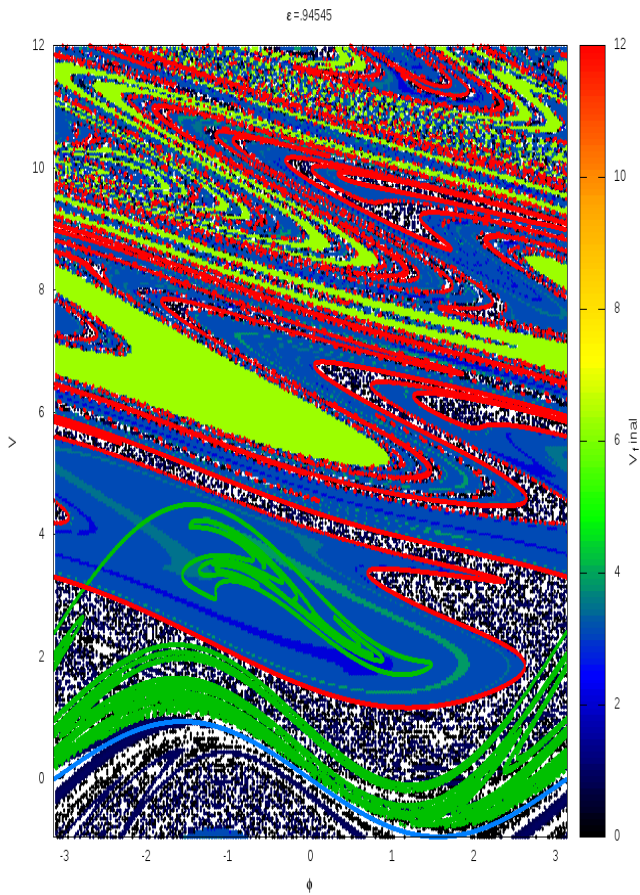}}
\end{center}
\caption{Color online: \it{Basin of attraction overlapped with the stable and unstable manifolds from the first sink, and the vibrating wall itself. 
One can notice that for $\epsilon\approx\epsilon_t$ and $\gamma=0.8$, the basin of attraction of the first sink is tangent to the growing 
branch of the stable manifold embedded with the chaotic bottom attractor, generating a tangent bifurcation inside the
attracting basin of the first sink. Yet, one can look at and see that from the place where the sink should be located, 
are rising three new branches, as a result from the tangent behaviour between the manifolds. The comparison with the basin of attraction
overlapped, shows better this behaviour.}}
\label{fig9}
\end{figure}

Still, we display the tangent bifurcation for the first sink in Fig.\ref{fig9}, where for $\epsilon\approx\epsilon_t$, we overlap 
the basin of attraction with the stable and unstable manifold from the first saddle and the vibrating wall itself. We can see that 
the basin of the first sink is tangent the growing branch of the bottom attractor. At the same time, we can see three branches
rising from the place where the sink should be located inside its basin of attraction. We also stress that if noise were added to the
dynamics \cite{ref28,ref46,ref47,ref48}, the scenario of the crises might would change, once the perturbation would be different. 
Of course, that noise should be consider in real experimental devices, so as a possible future work, we would be consider the
introduction of noise in the system.

\section{Final Remarks and Conclusions}
\label{sec4}
The dynamics of a bouncing ball model is investigated for a high dissipation regime. Some chaotic properties were set up, and average
properties of the velocity were obtained. Depending of the combination of control parameters many attractors can coexist, leading to a very
complex dynamics in the phase space. Basins of attraction for some attractors and their evolution with the increase of the perturbation 
parameter were characterized.  

Increasing the perturbation parameter leads us to characterize an unusual class of boundary crises, where the attractor collides 
physically with the vibrating wall. In these crises we observed a reduction in the number of attractors in the phase space.
These phenomena could be extended to other vibrating and impact systems with high dissipation regimes, as in human stability performance 
\cite{ref12} and microscopic vibration systems \cite{ref13}, in order to reduce the number of attractors (stable vibration modes) in such
systems. 

Also, we found the attractors are in an intertwined form. This was confirmed by the drawn of stable and unstable manifolds. Here, when a 
crises between manifolds occurs, it creates a successive destruction mechanism for all the attractors originated by the sinks, giving an
``extra life time" to the bottom chaotic attractor, turning it into a robust attractor. As a future work, it would be interesting to see 
with the bifurcation follows the Feigenbaum's $\delta$ relation, and if there would be any chance to find shrimp-like structures in the
parameter space \cite{ref49,ref50,ref51}. Also, we could consider the introduction of noise in the dynamics, to make a link with real
experimental devices.\

\acknowledgments
ALPL acknowledges CNPq and CAPES - Programa Ci\^encias sem Fronteiras - CsF
(0287-13-0) for financial support. IBL thanks FAPESP (2011/19296-1) and EDL thanks
FAPESP (2012/23688-5), CNPq and CAPES, Brazilian agencies. ALPL also thanks the University of
Bristol for the kindly hospitality during his stay in UK. This research was
supported by resources supplied by the Center for Scientific Computing
(NCC/GridUNESP) of the S\~ao Paulo State University (UNESP).

\end{document}